# Topological Phases of MnA$_2$X$_4$ (A=Bi, Sb; X = Se, Te) under Magnetic Field

Sugata Chowdhury[1,2,3*], Kevin F. Garrity[1], Francesca Tavazza[1]

[1]Materials Measurement Laboratory, National Institute for Standards and Technology, Gaithersburg, MD 20899; [2]Department of Physics and Astrophysics, Howard University, Washington DC, 20059; [3]IBM-Howard Quantum Center, Howard University, Washington DC, 20059; email: sugata.chowdhury@howard.edu

**Abstract:** The concept of electronic topology and the associated topological protection brings excellent opportunities for developing next-generation devices, especially those requiring minimal scattering or high quantum mechanical coherence. Ideally, magnetic topological materials (MTM) should have their Dirac/Weyl points and/or associated mass gaps at the Fermi energy ($E_F$) or be readily tunable such that they can be placed at $E_F$ via external perturbations such as electric field gating, chemical substitutions, or doping. Three-dimensional antiferromagnetic (AFM) materials like MnBi$_2$X$_4$ (X=Se, Te) that have strong spin-orbit coupling (SOC) and broken time-reversal symmetry (TRS) due to magnetic ordering have been the subject of enormous interest because they can display a variety of topological properties, including spin-polarized edge states even in the absence of an external magnetic field. In this work, using density functional theory (DFT), we have studied the electronic properties and topological phases of the first intrinsic magnetic topological insulator family MnBi$_2$Te$_4$ (MBT) in the presence of an external magnetic field. Our calculations reveal that the topological phase of bulk rhombohedral (R$\bar{3}$m) MnA$_2$X$_4$ (A = Sb, Bi; X=Se, Te) depends on the spin direction and the chemistry. The antiferromagnetic (AFM) ground state of MnSb$_2$Se$_4$ (MSS) is a trivial insulator, whereas the AFM ground state of MnBi$_2$Se$_4$ (MBS) is an Axion insulator. Both materials become nodal point or nodal line Weyl semimetals in the presence of a sufficiently strong external magnetic field. The AFM ground state of MnSb$_2$Te$_4$ (MST) is an Axion insulator. MST is a type-II Weyl semimetal with spins aligned in the $\hat{Z}$-direction, but becomes insulating with an inverted band gap for spins in-plane. Similarly, the AFM phase of MnBi$_2$Te$_4$ (MBT) is an Axion insulator, but remains insulating with an inverted gap in the ferromagnetic phase. Additionally, we demonstrated the evolution of the topological phase of MnBi$_2$Te$_4$ (MBT) by substituting the Bi atoms with the Sb atoms. This work indicates how to manipulate the topological and electronics phases of the MBT family via chemistry or external field.

**Introduction:** Materials with topologically protected electronic states display a variety of properties that are promising for next-generation electronic devices, specifically those requiring minimal scattering or high quantum mechanical coherence.[1-5] In particular, magnetic topological materials are necessary to enable the dissipationless currents of the quantum anomalous Hall effect.[1-4] Early progress in this area focused on doping magnetic atoms into known non-magnetic topological materials, but impurity-induced disorder ultimately limits the performance of this approach.[6-31] There is need for new materials that combine high magnetic ordering temperatures with robust topological properties.

Antiferromagnetic MnBi$_2$Te$_4$ (MBT)[1-3,32-76] has recently attracted enormous attention as an axion insulator in bulk samples and a quantum anomalous Hall insulator in thin films. MBT has many ideal properties for topological applications, including (1) bulk crystal stability, (2) a lack of trivial bands crossing the Fermi level, (3) topological features at the Fermi level, and (4) the ability to tune via electric field gating, external magnetic fields, or chemical substitution.[34,35,49,50,58,69] The family of materials related to MBT,[35,39,74,77] namely MnBi$_2$Se$_4$ (MBS),[42,44,64,73] MnSb$_2$Te$_4$ (MST),[33,67,78-82] MnSb$_2$Se$_4$ (MSS), are known to display some of the same properties as MBT, but they have not been as thoroughly investigated. In this work, we use first-principles calculations to systematically investigate the electronic structure of this materials family, comparing and rationalizing their ground state properties and the changes that occur under a magnetic field sufficient to align the spins.

MBT naturally forms a layered van der Waals (vdW) material with R-3m symmetry closely related to the topological insulator Bi$_2$Te$_3$, but with an extra layer of MnTe inserted into the quintuple layer, forming a septuple layer. MST is known to form a solid solution with MBT in the same crystal structure. Unlike MBT and MST, MBS's ground state crystal structure is an unrelated monoclinic phase (C2m symmetry), resulting in a topologically trivial semiconductor. However, the energy difference between that structure and the rhombohedral structure is small, and rhombohedral MBS has been successfully grown in thin films as either a single-phase or as a superlattice with Bi$_2$Se$_3$.[44,73] MBT, MST, and MBS are all known to have an AFM ground state spin ordering, with spins ferromagnetically aligned within layers. Surprisingly, very little has been reported experimentally or theoretically regarding the R-3m structure of MSS.

The magnetic and topological properties of MBT have been studied extensively. The magnetic ordering transition of MBT is around 25 K, and it undergoes a spin flop transition to a ferromagnetic phase under a magnetic field.[32,59,77] The quantum anomalous Hall effect (QAHE) has been observed in thin films near 1K.[59,77] There are several conflicting experimental reports on the surface band gap of MBT.[36,51,62,63,66,72-74,76] Earlier studies showed a gap opening below the Neel temparature[34,45,53,59,65,66,75,76] as expected due to time-reversal symmetry breaking on the surface, but several recent works show a gapless state both above and below the Neel temperature.[6,33,51,71] A recent scanning tunneling microscope (STM) study showed that the surface gap varies depending on the spatial location.[19,23,25,27,42,46,53,63,65,76] These results highlight the role of inhomogeneity in understanding the surface state, likely due to defects, magnetic domains, or spin disorder.[42] Sb doping into MBT has been used to modify the magnetic and electronic properties of

MBT.[14,20,33,49,50] Several experimental groups have successfully synthesized MnSb$_2$Te$_4$ (MST), a type-II Weyl semi-metal (WSM) in the presence of an external magnetic field.[33,39,50,51,69,79-82] The magnetic and electronic properties of MBS have received less experimental attention, but a Dirac cone has been observed using angle-resolved photoemission spectroscopy (ARPES).[73]

This study aims to systematically investigate and compare the electronic and magnetic properties of bulk MBS, MBT, MST, and MSS. First, we will discuss the ground state magnetic orderings and energetics. Second, we will discuss the effects of different spin orderings on the electronic structure, emphasizing those that might be reached experimentally by an external magnetic field. Third, we will examine the band structures as we artificially vary the strength of the spin-orbit coupling (SOC) in order to separate the influence of SOC from other chemical shifts. Finally, we will look at the changes in the band structure that occur as a fraction of Sb is substituted for Bi in MnBi$_{1-x}$Sb$_x$Se$_4$ and MnBi$_{1-x}$Sb$_x$Te$_4$.

**Results: Ground state structures**: First, we will examine the crystal symmetry, magnetic phases, and the direction of the magnetic moments of the ground state crystal structure for all the materials. Table 1 presents the computed energy difference between the rhombohedral (R$\bar{3}$m) and the monoclinic (C2m) structure presented in Table 1. Details of the crystal structures are depicted in the supplemental material (Fig. S1). We compared the energy difference between the bulk rhombohedral layered structure and the bulk monoclinic layered structure for all bulk AFM phase (out-of-plane) crystal structures. DFT calculations reveal that the bulk MBT and MST[39] crystallize in the rhombohedral structure. In contrast, MBS and MSS crystallize in the monoclinic structure but with only a small energy difference,[73] consistent with the successful growth of MBS in thin films. We will focus on the rhombohedral phase for the rest of this work.

**Table 1**: Energy difference between monoclinic and rhombohedral structures.

| Out-of-Plane AFM Phase | MBT | MBS | MST | MSS |
|---|---|---|---|---|
| Monoclinic (AFM) (meV/atom) | 26.87 | 0 | 24.51 | 0 |
| Rhombohedral (AFM) (meV/atom) | 0 | 2.06 | 0 | 1.98 |

Next, we will investigate the energies of four different magnetic structures: (a) out-of-plane AFM; (b) in-plane AFM; (c) out-of-plane FM, and (d) in-plane FM phases. Consistent with previous works, we find that the out-of-plane AFM phase is the ground state for all materials, with the calculated magnetic moment of ≈ 5$\mu_B$, consistent with Mn in the +2 state. Table 2 presents the energy difference between AFM and FM phase, the energy difference between in-plane and out-of-plane magnetic phases. The calculated bandgap for the AFM phase of the MBS, MBT MSS, and MST are 202 meV, 66 meV, 112 meV, and 59 meV, respectively. The calculated lattice parameters of MBS,[44,73] MBT,[59,70,71] and MST[69,70] agree with the experimental finding presented in Table S1.

**Table 2**: Band gap and energy difference between different magnetic structures of MBT family members. Top: Energy difference with respect to the ground state AFM out-of-plane phase, in meV/cell. Bottom: Topological phases. The bandgap (meV) of the ground state structure listed in parenthesis. OOP/IP refer to out-of-plane/in-plane magnetic moments.

| meV/cell | **MBT** | **MBS** | **MST** | **MSS** |
|---|---|---|---|---|
| $FM_{IP}$ - $AFM_{OOP}$ | 4.1 | 1.8 | 3.3 | 1.6 |
| $AFM_{IP}$ - $AFM_{OOP}$ | 1.5 | 0.9 | 1.2 | 0.8 |
| $FM_{OOP}$ - $AFM_{OOP}$ | 2.6 | 2.2 | 1.8 | 0.9 |
| **Topological Phase** | | | | |
| $AFM_{OOP}$ (bandgap meV) | Axion (66) | Axion (202) | Axion (59) | Trivial (112) |
| $AFM_{IP}$ | Axion | Axion | Axion | Trivial |
| $FM_{OOP}$ | Inverted Gap | Type-I WS | Type-II WS | Type-I WS |
| $FM_{IP}$ | Inverted Gap | Nodal line WS | Inverted Gap | Nodal line WS |

Despite having the same magnetic ground state, the differing chemistry of MBT, MBS, MST, and MSS have significant effects on the electronic structure and topology. Consistent with this work, previous first-principles studies have identified the ground state AFM phases of MBT[34-36,41,45,58,59,73,75] and MBS[64] as AFM topological insulators, a type of axion insulator. In contrast, there is no consensus regarding the topological phase of the AFM ground state of MST.[39,79-82] Our DFT calculations show that MST ground state has an inverted (non-trivial) band structure. Regardless, MST is close to the boundary between a trivial and inverted band structure and the exact phase may depend on the calculation approximations and crystal structure. In contrast to the other three materials, we find that the ground state AFM phase of MSS is a trivial semiconductor, with a gap of 112 meV. The bandgap of AFM-GS is mentioned in the parenthesis on the second part of Table 2. These results indicate that the MBT family's bandgap and topology depend strongly on the chemical composition. We will discuss all of these phases further in the following sections.

**Tuning the topological properties using an external magnetic field**: Next, we will consider the possibility of tuning the electronic and topological phases of MBT, MBS, MST, and MSS using an external magnetic field to alter the spin configuration. Along these lines, spin flop transitions to a ferromagnetic phase have been observed experimentally under magnetic fields perpendicular to the surface,[79,83,84] suggesting that such transitions should be possible in practice. Additionally, to make these phases easier to observe, it may be possible to induce ferromagnetic inter-layer magnetic couplings via doping, layering, or other mechanisms.[17,39,77,79]

Interestingly, every material shows differing topological behavior under the combination of the four main spin configurations we consider: out-of-plane AFM, in-plane (x-direction) AFM, out-of-plane FM, and in-plane FM. In addition to just considering the end states, we consider three magnetic transitions that could be followed by slowly rotating an external magnetic field, as shown in schematically in Fig. 1b. Fig. 2 shows in-plane AFM to out-of-plane FM; Fig. 3 shows out-of-plane AFM to in-plane FM, and Fig. 4 shows out-of-plane FM to in-plane FM. In each case, we show the evolution of the band structures of all four materials evolving from the starting spin configuration (left column) to the ending configuration (right column) in steps of 30°. The four rows correspond to MBS, MBT, MSS, and MST, respectively. Additional spin dependent band structures are presented in the SM.

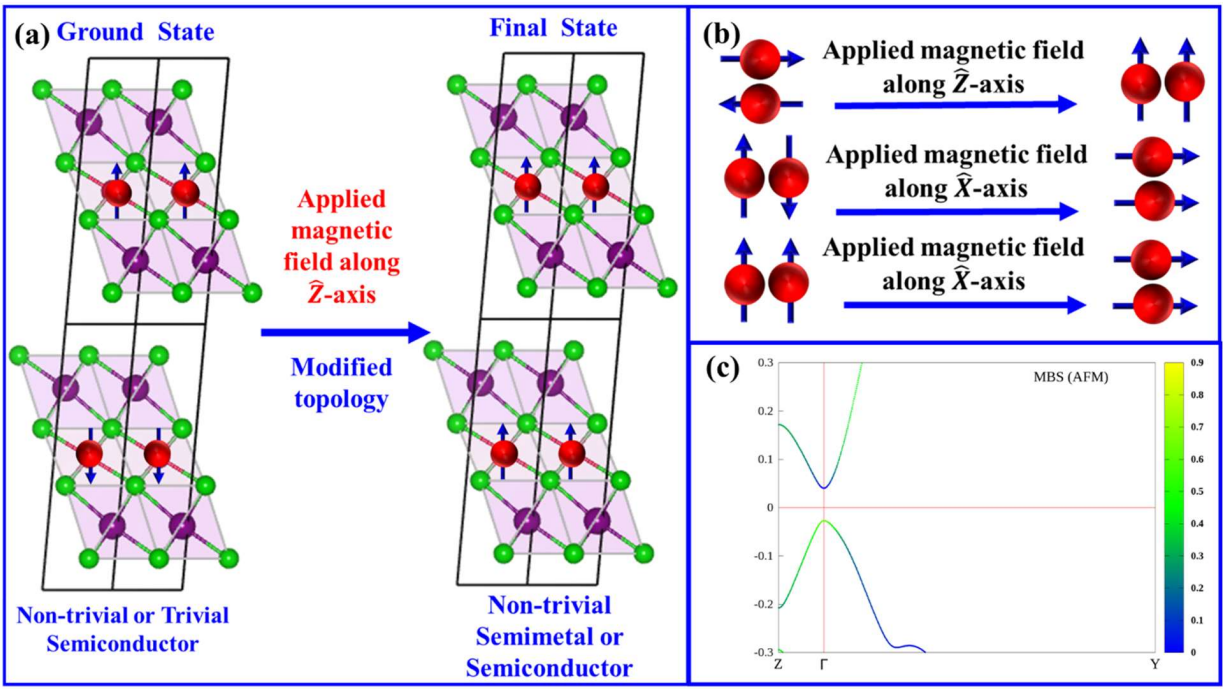

**Fig. 1:** (a) Side view of the initial and final bulk structure of $MnA_2X_4$ in absent and presents of the magnetic field, where the red, purple, and green ball represents Mn, heavy elements (Bi or Sb), and the green ball represents chalcogenides elements (Se or Te), respectively. (b) Three different magnetic moment directions of Mn atoms are considered for this work.

**In-plane AFM to out-of-plane FM**: Fig. 2 presents the evolution of the band structure as the spins are gradually rotated from an in-plane AFM to an out-of-plane FM configuration. These band structures are relevant to describing the canted ferromagnetic phase of MBT and MST after the spin flop transition occurs. In addition, they should be directly relevant to the observed in-plane AFM phase of MBS (and possibly

MSS). Although the crystal symmetry and chemical formula are the same for all the materials, there are significant differences in their bandstructures. All materials begin insulating in the AFM phase, although only MSS has a non-inverted band structure. MBS, MST, and MSS all evolve into semimetals under field along the z-direction, while MBT remains insulting with an inverted band structure. For MBS and MSS, the topological phase transition happens when the magnetic spin direction makes an angle greater than 40° with the $+\hat{X}$-axis. MST becomes a type-II Weyl semimetal due to the low energy conduction band that runs from Z to Γ, with a minimum in-between. In contrast, the band crossing in the Se compounds is type-I and much closer to Γ. The reduced band gaps and eventual crossings under field are driven by the loss of the combination of time-reversal symmetry plus a translation, which requires each band to be double degenerate in the AFM phase. Without this symmetry, the bands near the Fermi level become spin split, which is enough to close the gap except for MBT. MBT remains insulting with an inverted band structure, consistent with the experimental observation of the quantum anomalous Hall effect under the field.[36,82,83]

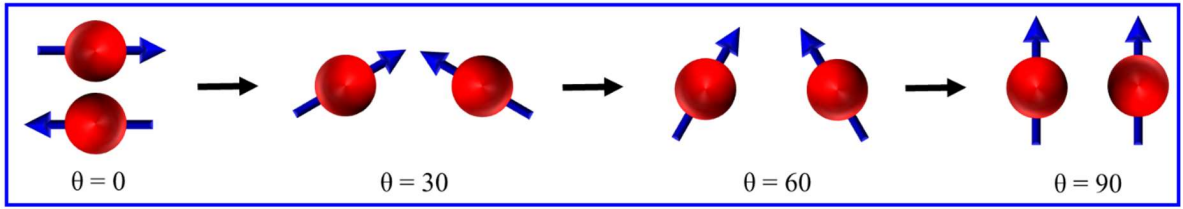

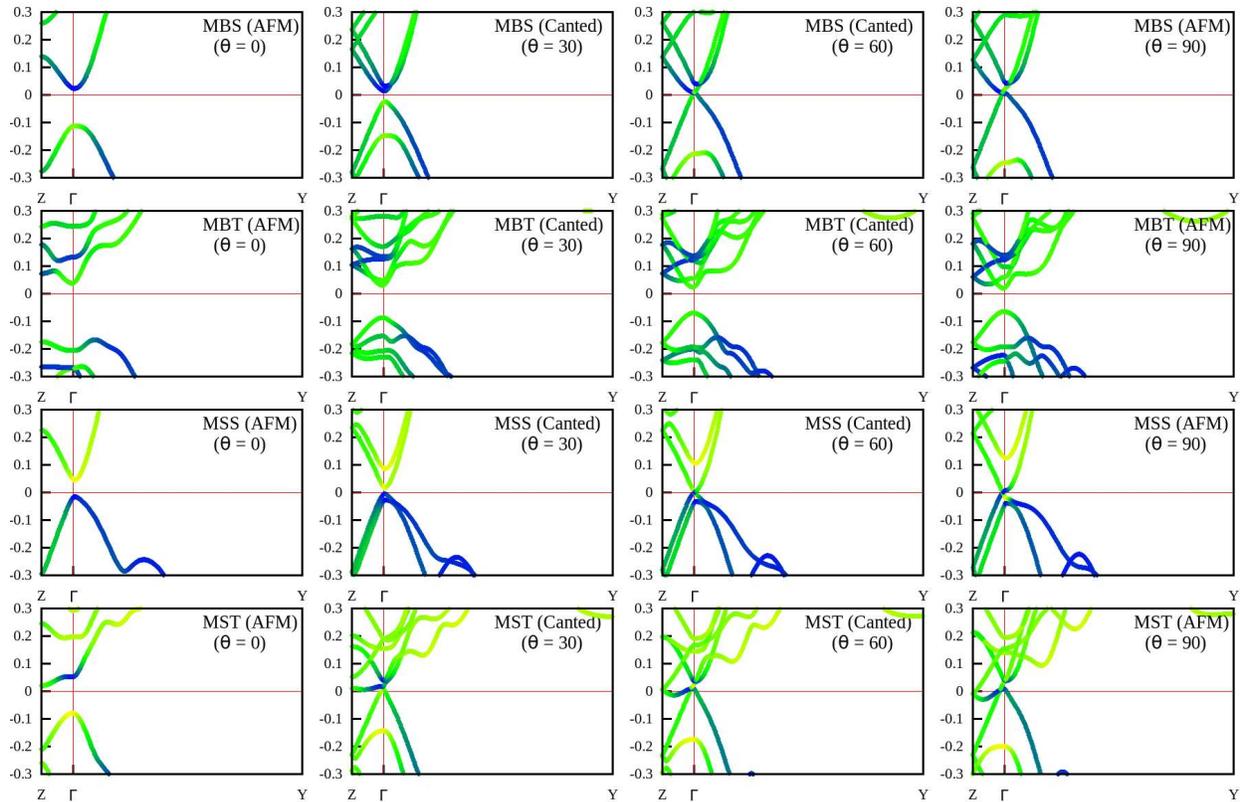

**Fig 2**: Evolution of the bulk bandstructure during the spin-flop transition from AFM in-plane (left column) to FM out-of-plane (right column). The direction of spin in the center panels is canted at 30 and 60 degrees. From top to bottom, materials are MBS, MBT, MSS, and MST. The color scale shows projection onto Bi-centered Wannier functions (see Fig. 1c).

The Weyl semimetal phase in MSS in the FM phase is notable because, unlike the other materials, it begins as a narrow gap trivial semiconductor in the AFM phase. To understand the differences between the materials in the FM phase, we have investigated the symmetry of the bands at the $\Gamma$ point near the Fermi level for each material, as summarized in Table 3. We find that none of the materials matches the trivial band ordering. The three semimetals all have the $\Gamma_{5+}$ band with Bi-like character occupied and $\Gamma_{4-}$ unoccupied due to band inversion. This results in a band crossing along the $\Gamma$-to-Z line when those bands cross with $\Gamma_4$ and $\Gamma_5$ symmetry. MBT instead has a double band inversion with both $\Gamma_{5+}$ and $\Gamma_{4+}$ occupied and $\Gamma_{5-}$ and $\Gamma_{4-}$ unoccupied. Because of the double band inversion, along the $\Gamma$-to-Z line there are two $\Gamma_4$ band and two $\Gamma_5$ bands each repel each other, keeping the gap open even though the atomic character of the occupied states changes along this line.

**Table 3**: Symmetries of the bands near the Fermi level at $\Gamma$ in the FM out-of-plane phase. Columns are second highest occupied (HO-1), HO, lowest unoccupied (LU), and LU+1.

| Materials (FM) | HO-1 | HO | LU | LU+1 |
|---|---|---|---|---|
| MBS | ;$\Gamma_{5+}$ | $\Gamma_{5-}$ | $\Gamma_{4-}$ | $\Gamma_{4+}$ |
| MBT | $\Gamma_{6+}$ | $\Gamma_{4+}$ | $\Gamma_{5-}$ | $\Gamma_{4-}$ |
| MSS | $\Gamma_{5+}$ | $\Gamma_{5-}$ | $\Gamma_{4-}$ | $\Gamma_{4+}$ |
| MST | $\Gamma_{5+}$ | $\Gamma_{5-}$ | $\Gamma_{4-}$ | $\Gamma_{4+}$ |
| Trivial | $\Gamma_{5-}$ | $\Gamma_{4-}$ | $\Gamma_{5+}$ | $\Gamma_{4+}$ |

**Out-of-plane AFM to in-plane FM**: The evolution of the band structures during the out-of-plane AFM to in-plane FM transitions is depicted in Fig. 3. Again, we find that each material is a semiconductor in the AFM phase, and only MSS has a non-inverted gap. As the spins rotate past 40° towards the X-direction, MSS and MBS become nodal line semimetals, consistent with previous calculations on MBS. The nodal line is in the Y-Z plane when the spins are in the X direction and is protected by a mirror symmetry. Despite the similar topology, the ordering of the bands at $\Gamma$ are again different in MSS and MBS, with two Bi-like states occupied in MBS and only one in MSS. In contrast to the Se-compounds, the band gaps of MBT and MST remain open during this transition. Unlike the previous transition, MST did not become a semimetal when the spins are aligned in-plane. The difference is that aligning the spins in the X direction raises the conduction band minimum along the Γ-Z line rather than lowering it. This difference prevents the bands from crossing at Γ even though the gap decreases at the Γ point, as will be discussed further in the following section.

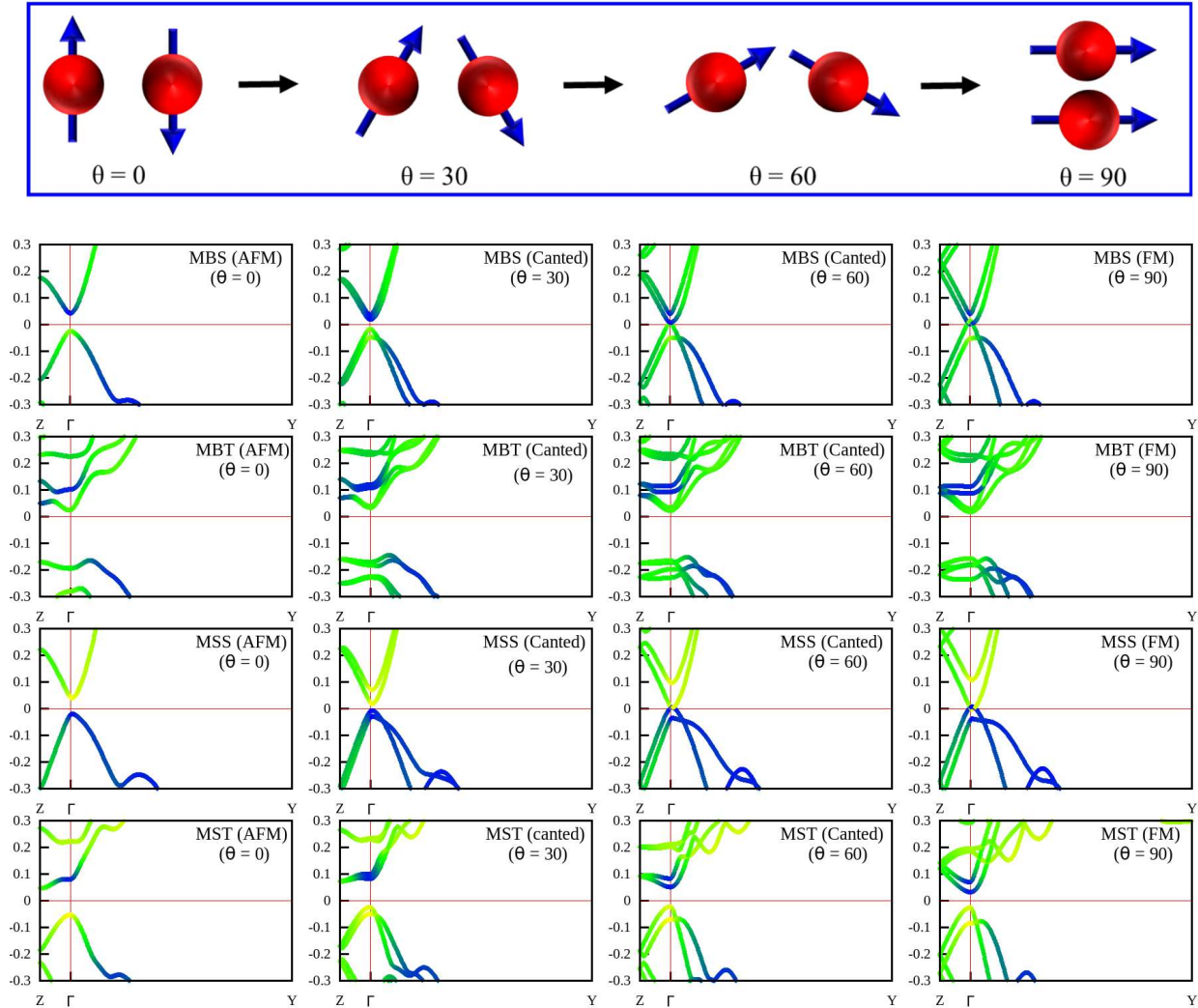

**Fig 3**: Evolution of the bulk bandstructure during the spin-flop transition from AFM out-of-plane (left column) to FM in-plane (right column). The direction of spin in the center panels is canted at 30 and 60 degrees. From top to bottom, materials are MBS, MBT, MSS, and MST. The color scale shows projection onto Bi-centered Wannier functions (see Fig. 1c).

**Out-of-plane FM to in-plane FM**: Finally, we will consider the evolution of the band structures during a transition between out-of-plane FM to in-plane FM (x-direction) phases, as shown in Fig. 4. As discussed earlier, the topological nature of the band structure of MBS depends on the direction of the magnetic field. Between 30° and 60° tilt, there is a band crossing at Γ that separates a semiconducting phase with a type-I

Weyl semimetal. This change is caused by a significant shift in the shape of the conduction band, with the minimum between Γ-Z disappearing as the spins are aligned more towards the X-direction. In the case of MBS and MSS, the systems remain semimetals throughout the transition. For a spin direction except along X, the system lacks the symmetry necessary to protect the Weyl nodal line, and instead, there are two Wely points. These points expand in the Y-Z plane and turn into the nodal line as the spin direction is tilted into the plane. In contrast, FM-MBT remains a semiconductor with an inverted band structure in all cases. The sensitivity of these electronic structures to the spin direction makes them an ideal playground to study and tune the properties of various types of magnetic semimetallic behavior.

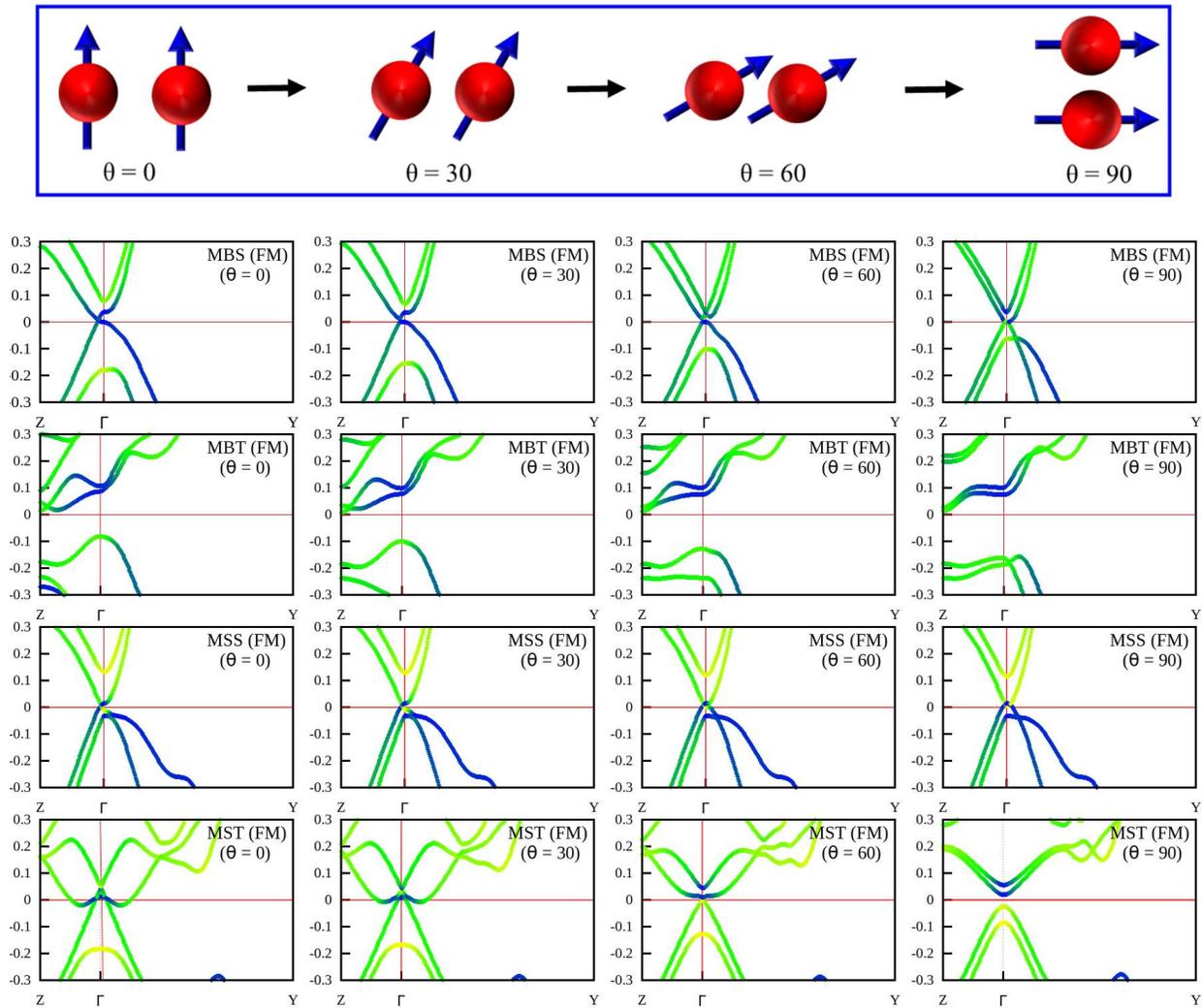

**Fig. 4**: Evolution of the bulk bandstructure during the transition from FM out-of-plane (left column) to FM in-plane (right column). The direction of spins in the center panels is canted at 30 and 60 degrees. From top to bottom, materials are MBS, MBT, MSS, and MST. The color scale shows projection onto Bi-centered Wannier functions (see Fig. 1c).

**Effect of spin-orbit coupling (SOC)**: SOC is the source of band inversion that leads to the various magnetic topological phases in this materials class. In order to understand the source of the differences between MBT, MBS, MST, and MSS, we artificially modify the strength of spin-orbit coupling in the out-of-plane FM phase, as shown in Fig. 5. Starting with the left column, we have set the SOC strength to 0, 0.25, 0.75, and 1 times the physical value. A similar plot for the AFM phase is presented in Fig. S3. As expected, all of the phases are topologically trivial semiconductors without SOC. Interestingly, all of the SOC=0 band shapes appear similar, although the gap sizes vary. The band gap of all of the materials begins decreasing as we increase the SOC, causing the materials to move towards band inversion at Γ. MBT has the largest change, while MSS changes least, which is expected given the larger SOC of Bi and Te versus Sb and Se.

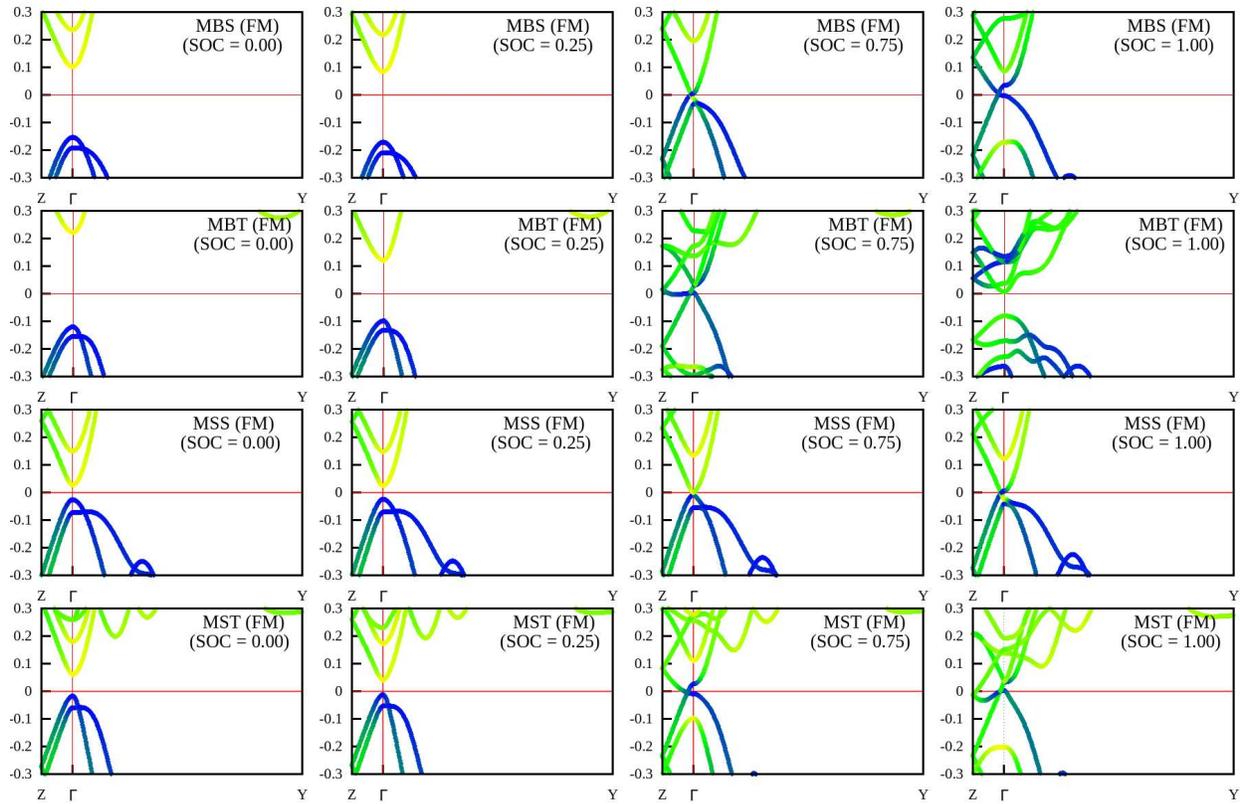

**Fig. 5.**: The evolution of the bulk bandstructure demonstrated the effect of the SOC on the topological phase of the MBT family. The top two rows of the fig represent $MnBi_2X_4$, and the bottom two rows represent $MnSb_2X_4$. The color bar shows projection onto Bi-centered Wannier functions.

As the SOC is increased in MBT, there are actually two phase transitions as different Bi bands become inverted. At 0.75, MBT is in a semimetallic phase due to the first band inversion, with the gap reopening at 1.0 due to the second band inversion. In addition to the band inversion at Γ, the conduction band minimum along Γ-Z decreases significantly towards the Fermi level as SOC increases. MST qualitatively follows a similar path but with less total change due to SOC. The physical 1.0 band structure of MST looks nearly identical to the semimetallic phase at SOC=0.75 of MBT. This suggests that the strong SOC of Bi+Te is key to the strong band inversion at Γ in MBT, while the shape of the conduction band from Γ-Z depends on the SOC of Te. MBS and MSS also change similarly due to SOC, although the changes are more limited in MSS. Again, the 0.75 band structure of MBS looks like the 1.0 band structure of MSS. The additional SOC of Bi relative to Sb results in a more robust band inversion in MBS relative to MSS, despite the larger SOC=0.0 gap of MBS. These plots suggest that differences in total SOC are the major factor in differences between the four phases. In addition, the SOC of Te is essential in changing the conduction band shape and causing MST to be a type-II Weyl semimetal while MBS and MSS are type-I.

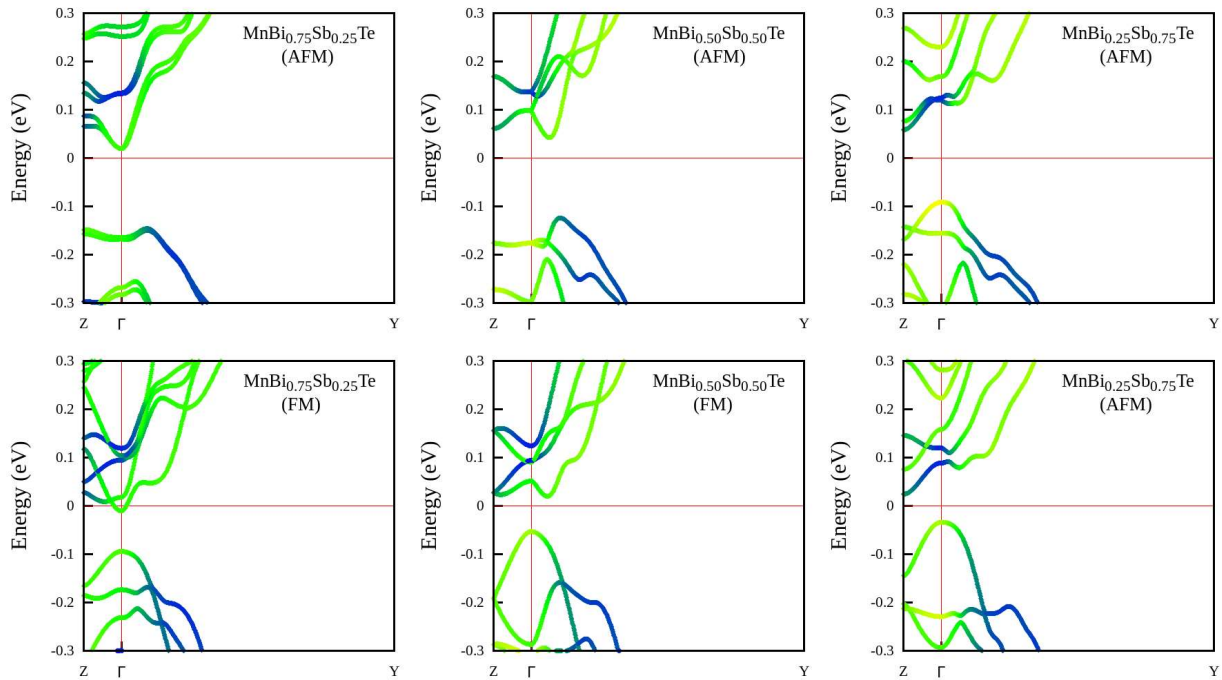

**Fig. 6**.: The evolution of the bulk bandstructure demonstrated the effect of the doping or chemical substitution of Bi atoms by Sb atoms on the out-of-plane AFM (top row) and out-of-plane FM phases of the MBT crystal. The left, center, and right columns are for $MnBi_{1.5}Sb_{0.5}Te_4$, $MnBi_{1.0}Sb_{1.0}Te_4$, and $MnBi_{0.5}Sb_{1.5}Te_4$, respectively. The color bar shows projection onto Bi-centered Wannier functions.

**Effect of chemical doping on MBT**: While we focus primarily on the end member phases in this work, our results suggest that alloying between these four materials could be a route to tune properties in interesting ways. As an example, we will consider the impact of doping Sb in place of Bi in MBT, as shown in Fig. 6. Here we considered 2×1×2 supercells and systematically replaced Bi atoms with Sb atoms to investigate three different Sb doped systems (25%, 50%, 75%). As shown in the top panels of Fig. 6, our calculations reveal no significant change in the AFM phase of MBT due to Sb doping, with the band gap remaining open and inverted in all cases. As discussed earlier, both MBT and MST are antiferromagnetic topological insulators in their ground states. Thus, doping Sb for Bi is an effective way to tune the SOC in this material. In contrast, there is a topological phase transition when replacing the Bi with Sb in the out-of-plane FM phase. The FM phase of MBT is insulating, whereas FM-MST is a type-II Weyl semimetal. As we increase the Sb concentration, the conduction band between Γ and Z begins to dip below the Fermi level. This ultimately results in the type-II Weyl semimetal phase when the concentration of the Sb atoms is more than 75%. These results are consistent with the Weyl semimetal phase observed experimentally in the MBT-MST system and demonstrate the importance of understanding all of the available knobs to tune the topological phases in septuple layer magnetic materials.[69,79,85]

**Conclusion**: In this work, we have systematically examined the effect of an external magnetic field, chemical substitution, and SOC on the topological phases and the electronic properties of the MBT family. In particular, we contrast the well-studied MBT with MBS, MSS, and MST. The ground state electronic structures of the MBT family are all out-of-plane AFM insulators, but they behave differently in the presence of an external magnetic field that can align the spins. Results indicate that the Se-based materials preferred C2m crystal symmetry, whereas the Te-based materials are thermodynamically form $R\bar{3}m$ crystal symmetry, which is a good agreement with previous reports. The Axion insulating phase of MBT does not change in the presence of an external magnetic field. The trivial AFM-MSS is transformed to a non-trivial type-I Weyl semimetal under any external magnetic field. In contrast, MST is a type-II Weyl semimetal when the magnetic field is along $\hat{Z}$-axis (out-of-plane), but it behaves as an Axion insulator when the magnetic field is along $\hat{X}$-axis (in-plane). The topological phase of the materials depends on the SOC strength of the materials. Furthermore, the topological phase of the materials also depends on the chemical substitutions and percent of Bi or Sb atoms. Overall, the methods underlying this work can be extended to describe and tunning the topological properties of other magnetic topological materials while also highlighting the effects that magnetic fields have on the topological phase and electronics properties of those materials systems.

**Computational:** Calculations were carried out using density-functional theory (DFT)[86,87] as implemented in the QUANTUM ESPRESSO code.[88] We used the PBEsol generalized gradient approximation as exchange and correlation potential.[89] We have used fully relativistic norm-conserving pseudopotentials.[90,91] We use a plane-wave cutoff of 70 Ry, a 10x10x6 Monkhorst-Pack[92] k-point mesh for bulk geometries, and a 8 × 8 × 1 k-point grid for slab geometries. For calculations with Mn, we use DFT+U[93,94] with U = 3 eV, although we find that our results are robust to changes in the value of U, as the states near the Fermi level are not predominantly Mn states. All the geometric structures are fully relaxed until the force on each atom is less than 0.002 eV/Å, and the energy-convergence criterion is $1 \times 10^{-6}$ eV. Results from our DFT calculations are then used as input to construct maximally localized Wannier functions using WANNIER90.[95,96]

**Ackonowledgement**: The work at Howard University was supported by the US Department of Energy (DOE), Office of Science, Basic Energy Sciences Grant No. DE-SC0022216 (modeling complex magnetic states in materials).

**Reference:**

1   Hasan, M. Z. & Kane, C. L. Colloquium: topological insulators. *Reviews of modern physics* **82**, 3045 (2010).
2   Kane, C. & Moore, J. Topological insulators. *Physics World* **24**, 32 (2011).
3   Qi, X.-L. & Zhang, S.-C. Topological insulators and superconductors. *Reviews of Modern Physics* **83**, 1057 (2011).
4   Haldane, F. D. M. Model for a quantum Hall effect without Landau levels: Condensed-matter realization of the" parity anomaly". *Physical review letters* **61**, 2015 (1988).
5   Essin, A. M., Moore, J. E. & Vanderbilt, D. Magnetoelectric polarizability and axion electrodynamics in crystalline insulators. *Physical review letters* **102**, 146805 (2009).
6   Chang, C.-Z. *et al.* Experimental observation of the quantum anomalous Hall effect in a magnetic topological insulator. *Science* **340**, 167-170 (2013).
7   Checkelsky, J. *et al.* Trajectory of the anomalous Hall effect towards the quantized state in a ferromagnetic topological insulator. *Nature Physics* **10**, 731-736 (2014).
8   Kou, X. *et al.* Scale-invariant quantum anomalous Hall effect in magnetic topological insulators beyond the two-dimensional limit. *Physical review letters* **113**, 137201 (2014).
9   Bestwick, A. *et al.* Precise quantization of the anomalous Hall effect near zero magnetic field. *Physical review letters* **114**, 187201 (2015).
10  Chang, C.-Z. *et al.* High-precision realization of robust quantum anomalous Hall state in a hard ferromagnetic topological insulator. *Nature materials* **14**, 473-477 (2015).
11  Chen, Y. *et al.* Massive Dirac fermion on the surface of a magnetically doped topological insulator. *Science* **329**, 659-662 (2010).
12  Xu, S.-Y. *et al.* Hedgehog spin texture and Berry's phase tuning in a magnetic topological insulator. *Nature Physics* **8**, 616-622 (2012).
13  Rienks, E. *et al.* Large magnetic gap at the Dirac point in a Mn-induced Bi $_2$ Te $_3$ heterostructure. *arXiv preprint arXiv:1810.06238* (2018).
14  Zhang, J.-M. *et al.* Stability, electronic, and magnetic properties of the magnetically doped topological insulators Bi 2 Se 3, Bi 2 Te 3, and Sb 2 Te 3. *Physical Review B* **88**, 235131 (2013).
15  Abdalla, L., Seixas, L., Schmidt, T., Miwa, R. & Fazzio, A. Topological insulator Bi 2 Se 3 (111) surface doped with transition metals: An ab initio investigation. *Physical Review B* **88**, 045312 (2013).
16  Růžička, J. *et al.* Structural and electronic properties of manganese-doped Bi2Te3 epitaxial layers. *New Journal of Physics* **17**, 013028 (2015).

17    Lee, J. S. *et al.* Ferromagnetism and spin-dependent transport in n-type Mn-doped bismuth telluride thin films. *Physical Review B* **89**, 174425 (2014).
18    Figueroa, A. I. *et al.* Local structure and bonding of transition metal dopants in Bi2Se3 topological insulator thin films. *The Journal of Physical Chemistry C* **119**, 17344-17351 (2015).
19    Zhang, D. *et al.* Interplay between ferromagnetism, surface states, and quantum corrections in a magnetically doped topological insulator. *Physical Review B* **86**, 205127 (2012).
20    Choi, J. *et al.* Magnetic properties of Mn-doped Bi2Te3 and Sb2Te3. *physica status solidi (b)* **241**, 1541-1544 (2004).
21    Janíček, P., Drašar, Č., Lošť'Ák, P., Vejpravová, J. & Sechovský, V. Transport, magnetic, optical and thermodynamic properties of Bi2− xMnxSe3 single crystals. *Physica B: Condensed Matter* **403**, 3553-3558 (2008).
22    Hor, Y. *et al.* Development of ferromagnetism in the doped topological insulator Bi 2− x Mn x Te 3. *Physical Review B* **81**, 195203 (2010).
23    Liu, Q., Liu, C.-X., Xu, C., Qi, X.-L. & Zhang, S.-C. Magnetic impurities on the surface of a topological insulator. *Physical review letters* **102**, 156603 (2009).
24    Sessi, P. *et al.* Signatures of Dirac fermion-mediated magnetic order. *Nature communications* **5**, 1-8 (2014).
25    Zhu, J.-J., Yao, D.-X., Zhang, S.-C. & Chang, K. Electrically controllable surface magnetism on the surface of topological insulators. *Physical review letters* **106**, 097201 (2011).
26    Sánchez-Barriga, J. *et al.* Nonmagnetic band gap at the Dirac point of the magnetic topological insulator (Bi 1− x Mn x) 2 Se 3. *Nature communications* **7**, 1-10 (2016).
27    Rosenberg, G. & Franz, M. Surface magnetic ordering in topological insulators with bulk magnetic dopants. *Physical Review B* **85**, 195119 (2012).
28    Ado, I., Dmitriev, I., Ostrovsky, P. & Titov, M. Anomalous Hall effect with massive Dirac fermions. *EPL (Europhysics Letters)* **111**, 37004 (2015).
29    Zhang, J. *et al.* Topology-driven magnetic quantum phase transition in topological insulators. *Science* **339**, 1582-1586 (2013).
30    Zhang, Z. *et al.* Electrically tuned magnetic order and magnetoresistance in a topological insulator. *Nature communications* **5**, 1-7 (2014).
31    Liu, M. *et al.* Crossover between weak antilocalization and weak localization in a magnetically doped topological insulator. *Physical review letters* **108**, 036805 (2012).
32    Aliev, Z. S. *et al.* Novel ternary layered manganese bismuth tellurides of the MnTe-Bi2Te3 system: Synthesis and crystal structure. *Journal of Alloys and Compounds* **789**, 443-450 (2019).
33    Chen, B. *et al.* Intrinsic magnetic topological insulator phases in the Sb doped MnBi 2 Te 4 bulks and thin flakes. *Nature communications* **10**, 1-8 (2019).
34    Chen, Y. *et al.* Topological electronic structure and its temperature evolution in antiferromagnetic topological insulator MnBi 2 Te 4. *Physical Review X* **9**, 041040 (2019).
35    Cho, Y. *et al.* Phonon modes and Raman signatures of MnBi2nTe3n+ 1 (n= 1, 2, 3, 4) magnetic topological heterostructures. *arXiv preprint arXiv:2107.03204* (2021).
36    Deng, Y. *et al.* Quantum anomalous Hall effect in intrinsic magnetic topological insulator MnBi2Te4. *Science* **367**, 895-900 (2020).
37    Ding, L. *et al.* Crystal and magnetic structures of magnetic topological insulators MnBi 2 Te 4 and MnBi 4 Te 7. *Physical Review B* **101**, 020412 (2020).
38    Eremeev, S., Otrokov, M. & Chulkov, E. V. Competing rhombohedral and monoclinic crystal structures in MnPn2Ch4 compounds: An ab-initio study. *Journal of Alloys and Compounds* **709**, 172-178 (2017).
39    Eremeev, S. *et al.* Topological magnetic materials of the (MnSb2Te4)·(Sb2Te3) n van der Waals compounds family. *The Journal of Physical Chemistry Letters* **12**, 4268-4277 (2021).

40  Eremeev, S. V., Otrokov, M. M. & Chulkov, E. V. New universal type of interface in the magnetic insulator/topological insulator heterostructures. *Nano letters* **18**, 6521-6529 (2018).
41  Estyunin, D. *et al.* Signatures of temperature driven antiferromagnetic transition in the electronic structure of topological insulator MnBi2Te4. *APL Materials* **8**, 021105 (2020).
42  Garrity, K. F., Chowdhury, S. & Tavazza, F. M. Topological surface states of Mn Bi 2 Te 4 at finite temperatures and at domain walls. *Physical Review Materials* **5**, 024207 (2021).
43  Gong, Y. *et al.* Experimental realization of an intrinsic magnetic topological insulator. *Chinese Physics Letters* **36**, 076801 (2019).
44  Hagmann, J. A. *et al.* Molecular beam epitaxy growth and structure of self-assembled Bi2Se3/Bi2MnSe4 multilayer heterostructures. *New Journal of Physics* **19**, 085002 (2017).
45  Hao, Y.-J. *et al.* Gapless surface Dirac cone in antiferromagnetic topological insulator MnBi 2 Te 4. *Physical Review X* **9**, 041038 (2019).
46  Hirahara, T. *et al.* Large-gap magnetic topological heterostructure formed by subsurface incorporation of a ferromagnetic layer. *Nano letters* **17**, 3493-3500 (2017).
47  Hirahara, T. *et al.* Fabrication of a novel magnetic topological heterostructure and temperature evolution of its massive Dirac cone. *Nature communications* **11**, 1-8 (2020).
48  Hu, C. *et al.* A van der Waals antiferromagnetic topological insulator with weak interlayer magnetic coupling. *Nature communications* **11**, 1-8 (2020).
49  Hu, C. *et al.* Tuning magnetism and band topology through antisite defects in Sb-doped MnBi 4 Te 7. *Physical Review B* **104**, 054422 (2021).
50  Hu, C., Tanatar, M. A., Prozorov, R. & Ni, N. Unusual dynamic susceptibility arising from soft ferromagnetic domains in MnBi8Te13 and Sb-doped MnBi2nTe3n+ 1 (n= 2, 3). *arXiv preprint arXiv:2106.08969* (2021).
51  Lee, S. H. *et al.* Spin scattering and noncollinear spin structure-induced intrinsic anomalous Hall effect in antiferromagnetic topological insulator MnB i 2 T e 4. *Physical Review Research* **1**, 012011 (2019).
52  Li, B. *et al.* Competing magnetic interactions in the antiferromagnetic topological insulator MnBi 2 Te 4. *Physical review letters* **124**, 167204 (2020).
53  Li, H. *et al.* Dirac surface states in intrinsic magnetic topological insulators EuSn 2 As 2 and MnBi 2 n Te 3 n+ 1. *Physical Review X* **9**, 041039 (2019).
54  Li, J. *et al.* Intrinsic magnetic topological insulators in van der Waals layered MnBi2Te4-family materials. *Science Advances* **5**, eaaw5685 (2019).
55  Lian, B., Liu, Z., Zhang, Y. & Wang, J. Flat chern band from twisted bilayer mnbi 2 te 4. *Physical review letters* **124**, 126402 (2020).
56  Liu, C. *et al.* Robust axion insulator and Chern insulator phases in a two-dimensional antiferromagnetic topological insulator. *Nature materials* **19**, 522-527 (2020).
57  Otrokov, M. *et al.* Magnetic extension as an efficient method for realizing the quantum anomalous hall state in topological insulators. *JETP Letters* **105**, 297-302 (2017).
58  Otrokov, M. *et al.* Unique thickness-dependent properties of the van der Waals interlayer antiferromagnet MnBi 2 Te 4 films. *Physical review letters* **122**, 107202 (2019).
59  Otrokov, M. M. *et al.* Prediction and observation of an antiferromagnetic topological insulator. *Nature* **576**, 416-422 (2019).
60  Otrokov, M. M. *et al.* Highly-ordered wide bandgap materials for quantized anomalous Hall and magnetoelectric effects. *2D Materials* **4**, 025082 (2017).
61  Peng, Y. & Xu, Y. Proximity-induced Majorana hinge modes in antiferromagnetic topological insulators. *Physical Review B* **99**, 195431 (2019).
62  Rienks, E. D. *et al.* Large magnetic gap at the Dirac point in Bi 2 Te 3/MnBi 2 Te 4 heterostructures. *Nature* **576**, 423-428 (2019).

63  Shikin, A. M. *et al.* Nature of the Dirac gap modulation and surface magnetic interaction in axion antiferromagnetic topological insulator $\mathrm{MnBi}_2\mathrm{Te}_4$ MnBi 2 Te 4. *Scientific Reports* **10**, 1-13 (2020).
64  Sugata, C., Garrity, K. F. & Francesca, T. Prediction of Weyl semimetal and antiferromagnetic topological insulator phases in Bi 2 MnSe 4. *NPJ Computational Materials* **5** (2019).
65  Swatek, P. *et al.* Gapless Dirac surface states in the antiferromagnetic topological insulator MnBi 2 Te 4. *Physical Review B* **101**, 161109 (2020).
66  Vidal, R. *et al.* Surface states and Rashba-type spin polarization in antiferromagnetic MnBi 2 Te 4 (0001). *Physical Review B* **100**, 121104 (2019).
67  Wu, J. *et al.* Natural van der Waals heterostructural single crystals with both magnetic and topological properties. *Science advances* **5**, eaax9989 (2019).
68  Xu, B. *et al.* Infrared study of the multiband low-energy excitations of the topological antiferromagnet MnBi 2 Te 4. *Physical Review B* **103**, L121103 (2021).
69  Yan, J.-Q. *et al.* Evolution of structural, magnetic, and transport properties in MnBi 2− x Sb x Te 4. *Physical Review B* **100**, 104409 (2019).
70  Yan, J.-Q. *et al.* Crystal growth and magnetic structure of MnBi 2 Te 4. *Physical Review Materials* **3**, 064202 (2019).
71  Zeugner, A. *et al.* Chemical aspects of the candidate antiferromagnetic topological insulator MnBi2Te4. *Chemistry of Materials* **31**, 2795-2806 (2019).
72  Zhang, D. *et al.* Topological axion states in the magnetic insulator MnBi 2 Te 4 with the quantized magnetoelectric effect. *Physical review letters* **122**, 206401 (2019).
73  Zhu, T. *et al.* Synthesis, Magnetic Properties, and Electronic Structure of Magnetic Topological Insulator MnBi2Se4. *Nano Letters* (2021).
74  Zhang, R.-X., Wu, F. & Sarma, S. D. Möbius insulator and higher-order topology in MnBi 2 n Te 3 n+ 1. *Physical review letters* **124**, 136407 (2020).
75  Vidal, R. C. *et al.* Topological electronic structure and intrinsic magnetization in MnBi 4 Te 7: A Bi 2 Te 3 derivative with a periodic Mn sublattice. *Physical Review X* **9**, 041065 (2019).
76  Nevola, D. *et al.* Coexistence of surface ferromagnetism and a gapless topological state in MnBi 2 Te 4. *Physical Review Letters* **125**, 117205 (2020).
77  Klimovskikh, I. I. *et al.* Tunable 3D/2D magnetism in the (MnBi 2 Te 4)(Bi 2 Te 3) m topological insulators family. *npj Quantum Materials* **5**, 1-9 (2020).
78  Chen, Y. *et al.* Ferromagnetism in van der Waals compound MnS b 1.8 B i 0.2 T e 4. *Physical Review Materials* **4**, 064411 (2020).
79  Lee, S. H. *et al.* Evidence for a Magnetic-Field-Induced Ideal Type-II Weyl State in Antiferromagnetic Topological Insulator Mn (Bi 1− x Sb x ) 2 Te 4. *Physical Review X* **11**, 031032 (2021).
80  Wimmer, S. *et al.* Ferromagnetic MnSb2Te4: A topological insulator with magnetic gap closing at high Curie temperatures of 45-50 K. *arXiv preprint arXiv:2011.07052* (2020).
81  Zhou, L. *et al.* Topological phase transition in the layered magnetic compound MnSb 2 Te 4: Spin-orbit coupling and interlayer coupling dependence. *Physical Review B* **102**, 085114 (2020).
82  Shi, G. *et al.* Anomalous Hall effect in layered ferrimagnet MnSb2Te4. *Chinese Physics Letters* **37**, 047301 (2020).
83  Bac, S.-K. *et al.* Topological response of the anomalous Hall effect in MnBi2Te4 due to magnetic canting. *arXiv preprint arXiv:2103.15801* (2021).
84  Liu, C. *et al.* Magnetic-field-induced robust zero Hall plateau state in MnBi2Te4 Chern insulator. *Nature Communications* **12**, 1-8 (2021).
85  Eremeev, S. V. *et al.* Topological magnetic materials of the (MnSb2Te4)·(Sb2Te3) n van der Waals compounds family. *The Journal of Physical Chemistry Letters* **12**, 4268-4277 (2021).

86 Hohenberg, P. & Kohn, W. Inhomogeneous electron gas. *Physical review* **136**, B864 (1964).
87 Kohn, W. & Sham, L. J. Self-consistent equations including exchange and correlation effects. *Physical review* **140**, A1133 (1965).
88 Giannozzi, P. *et al.* QUANTUM ESPRESSO: a modular and open-source software project for quantum simulations of materials. *Journal of physics: Condensed matter* **21**, 395502 (2009).
89 Perdew, J. P. *et al.* Restoring the density-gradient expansion for exchange in solids and surfaces. *Physical Review Letters* **100**, 136406 (2008).
90 Hamann, D. Optimized norm-conserving Vanderbilt pseudopotentials. *Physical Review B* **88**, 085117 (2013).
91 Schlipf, M. & Gygi, F. Optimization algorithm for the generation of ONCV pseudopotentials. *Computer Physics Communications* **196**, 36-44 (2015).
92 Monkhorst, H. J. & Pack, J. D. Special points for Brillouin-zone integrations. *Physical review B* **13**, 5188 (1976).
93 Liechtenstein, A., Anisimov, V. & Zaanen, J. Density-functional theory and strong interactions: Orbital ordering in Mott-Hubbard insulators. *Physical Review B* **52**, R5467 (1995).
94 Dudarev, S., Botton, G., Savrasov, S., Humphreys, C. & Sutton, A. Electron-energy-loss spectra and the structural stability of nickel oxide: An LSDA+ U study. *Physical Review B* **57**, 1505 (1998).
95 Marzari, N. & Vanderbilt, D. Maximally localized generalized Wannier functions for composite energy bands. *Physical review B* **56**, 12847 (1997).
96 Mostofi, A. A. *et al.* wannier90: A tool for obtaining maximally-localised Wannier functions. *Computer physics communications* **178**, 685-699 (2008).